\documentclass[10pt,pra,superscriptaddress,twocolumn,longbibliography]{revtex4-2}

\usepackage[T1]{fontenc}
\usepackage{amsmath}
\usepackage{amssymb}
\usepackage{graphicx}
\usepackage{siunitx}
\usepackage{datetime}
\usepackage{color}
\usepackage{newtxtext,newtxmath}

\usepackage{microtype}
\usepackage{braket}
\usepackage{hyperref}
\hypersetup{%
    bookmarksopen=false,
    bookmarksnumbered=true,
    pdfnewwindow=true,
    unicode=false,
    colorlinks=true,%
    citecolor=blue,
    linkcolor=black,
    urlcolor=blue,
    filecolor=blue
    }
    
\newcommand{\vb}[1]{\boldsymbol{#1}}
\newcommand{\rmi}{\mathrm{i}}
\DeclareMathOperator{\tr}{tr}

\begin{document}
\title{Entanglement structures in disordered chains of nitrogen-vacancy centers}
\author{Alexander M.~Minke}
\affiliation{Physikalisches Institut, Albert-Ludwigs-Universit\"{a}t Freiburg, Hermann-Herder-Stra{\ss}e 3, D-79104, Freiburg, Federal Republic of Germany}
\author{Andreas Buchleitner}
\email[]{a.buchleitner@physik.uni-freiburg.de}
\affiliation{Physikalisches Institut, Albert-Ludwigs-Universit\"{a}t Freiburg, Hermann-Herder-Stra{\ss}e 3, D-79104, Freiburg, Federal Republic of Germany}
\affiliation{EUCOR Centre for Quantum Science and Quantum Computing, Albert-Ludwigs-Universit\"{a}t Freiburg, Hermann-Herder-Stra{\ss}e 3, D-79104, Freiburg, Federal Republic of Germany}
\author{Edoardo G.~Carnio}
\email[]{edoardo.carnio@physik.uni-freiburg.de}
\affiliation{Physikalisches Institut, Albert-Ludwigs-Universit\"{a}t Freiburg, Hermann-Herder-Stra{\ss}e 3, D-79104, Freiburg, Federal Republic of Germany}
\affiliation{EUCOR Centre for Quantum Science and Quantum Computing, Albert-Ludwigs-Universit\"{a}t Freiburg, Hermann-Herder-Stra{\ss}e 3, D-79104, Freiburg, Federal Republic of Germany}

\date{\today, \currenttime}
\begin{abstract}
	A recent study [Phys.~Rev.~B \textbf{106} 174111 (2022)] has hypothesized the assembly, along a specific type of one-dimensional defects of diamond, of chains of nitrogen-vacancy (NV) centers, potentially enabling the creation of qubit registers via their dipole-coupled electron spins.
	Here we investigate the connectivity of chains of up to ten coupled spins, mediated by the bi- and multipartite entanglement of their eigenstates.
	Rather conveniently, for regularly spaced spins the vast majority of the eigenstates displays strong connectivity, especially towards the center of the spectrum and for longer chains. Furthermore, positional disorder can change, and possibly reduce, the connectivity of the register, but seldom suppresses it.
\end{abstract}
\maketitle
\section{Introduction}

Nitrogen-vacancy (NV) centers in diamonds are promising candidates for the realization of quantum computation platforms \cite{2018NaPho..12..516A, Chatterjee2021} since, thanks to the optical addressability of their electron and nuclear spins, direct state initialization, manipulation and readout \cite{7478018, PhysRevA.97.062330, doi:10.1126/science.1181193, PhysRevLett.119.223602} are possible in tandem with remarkably long coherence times \cite{Herbschleb2019, Abobeih2018}.
By coupling the electron spin of one NV center to nearby nuclear spins \cite{Fuchs2011,Morton2008,Childress2006}
it has been possible to control up to 27 qubits \cite{Abobeih2019}, and to entangle ten \cite{Bradley_2019}.
Quantum registers built on individual NV centers, however, offer limited scalability \cite{DiVincenzo_2000}, as spins further away become a source of decoherence \cite{Childress2006}.
Consequently, several mechanisms have been proposed to couple multiple NV centers:
optically 
\cite{Moehring2007,Togan2010}, by means of mechanical resonators \cite{Albrecht_2013, Golter2016} or flux qubits \cite{PhysRevLett.92.247902, PhysRevLett.105.210501}, or by magnetic dipolar interaction \cite{Bermudez2011, Yao_2012}.
The latter, in particular, requires closely spaced, and therefore strongly coupled, NV centers, to carry out operations within the coherence time of the involved spins \cite{10.1063/5.0007444}. However, the formation of defects at predetermined locations, typically by ion implantation \cite{Smith2019}, still poses a significant technological challenge \cite{PhysRevLett.123.106802, Luehmann2019}. So far, samples with only two \cite{Dolde2013,Dolde_2014,Neumann_2008,Neumann_2010} or three \cite{Haruyama_2019} coupled NV centers have been fabricated, with a creation yield decreasing by an order of magnitude for each additional NV center \cite{Haruyama_2019}.
Alternatively to implantation, a recent density functional theory analysis \cite{PhysRevB.106.174111} showed that, to minimize the formation energy, an NV center should align along the core of a $\SI{30}{\degree}$ partial dislocation (a one-dimensional defect in diamond), favoring the assembly of a chain of coupled NV centers.
Can one build a connected qubit register out of such a chain?
To answer this question, we study the entanglement in the eigenstates that emerge by all-to-all dipolar interactions between the NV centers.

Entanglement is the vital resource in all applications of quantum information, such as quantum computing, cryptography and teleportation \cite{RevModPhys.81.865, RevModPhys.74.197}.
In particular, certain classes of entangled states are useful for specific quantum information applications \cite{hein2006entanglement,GUHNE20091}. Still, the classification of entangled states of many-body (``multipartite'') systems is an ongoing field of research, and already a difficult task for more than three particles \cite{Mintert:2005aa,GUHNE20091,Bengtsson_2017,Friis_2018,Benatti_2020}

In this article, we devise a custom scheme 
to characterize the different entanglement structures, based on the bipartite or multipartite entanglement encoded in the eigenstates.
These structures enable the transmission of information across the chain \cite{Scholak_2011, Levi2015, Bradley_2019},
realizing, out of the electronic spins, a quantum \emph{bus} that shuttles information between the long-lived nuclear spins of the NV centers. This synergy between a ``nuclear memory'' and an ``electronic bus'' has already been demonstrated experimentally for two dipole-coupled NV centers \cite{Dolde2013,Dolde_2014}, and mirrors how trapped-ion quantum computers work, where the phonons of the (shared) ion motional modes provide the bus to the ion qubits \cite{Bruzewicz2019}.

The entanglement structures appearing in the spectrum of the chain convey the connectivity of the (electronic) qubit register, i.e., they reveal which eigenstates can be exploited -- e.g., in optimal quantum control schemes \cite{Dolde_2014,Lindel2023,Maile2024} -- to implement the desired dynamics (like a two-qubit gate) involving specific qubits.
We target, in particular, the resilience of these entanglement structures against geometric disorder, which is unavoidable in the random assembly of the NV centers along the dislocation.
Note that we here study the entire eigenspace of finite spin chains.
In the fabrication of quantum computing devices, any infinite-chain or thermodynamic limit is irrelevant, as either producer or user will only address finitely many components at a time.

In terms of organization of the content, the reader will find in Sec.~\ref{sec:chains} the model of the chains of coupled NV centers, in Sec.~\ref{sec:ent-quantification} a compact introduction to the entropy of entanglement and to Wootters's concurrence, which are employed in our classification scheme, described in Sec.~\ref{sec:automated-classification} together with the specific structures that we aim to recognize. In Sec.~\ref{sec:discussion} we apply the classification scheme to the spectra of regularly spaced chains, and then discuss how the observed entanglement structures change with increasing disorder strength. We finally draw our conclusions in Sec.~\ref{sec:conclusions}.

\section{Model and methodology}
\subsection{Chains of NV centers}\label{sec:chains}
The nitrogen-vacancy center shows trigonal symmetry around the axis connecting the nitrogen and the vacant carbon atom \cite{Manson2006}. In this work, the NV centers are placed on, and their axes aligned to, a dislocation \cite{PhysRevB.106.174111} that defines the $z$ axis. We take each NV center to be exposed to a local magnetic field.

The electronic degrees of freedom of a single negatively charged NV center, labelled $j$, are described by a spin-1 particle \cite{Manson2006} with spin operators $\vb S_j$ and Hamiltonian (now and always in units of $\hbar$) \cite{Bermudez2011,Yao_2012}
\begin{equation}\label{eq:ZFS-ham}
	H_j^{(e)} = D (S_j^z)^2 + g_e \mu_B \vb B_j \cdot \vb S_j ,
\end{equation}
where $D = \SI{2.87}{GHz}$ \cite{Loubser_1978} is the magnitude of the zero-field splitting, $g_e$ is the electron $g$-factor, $\mu_B$ the Bohr magneton, and $g_e\mu_B = \SI{2.8}{MHz/G}$ \cite{Loubser_1978}. If we write the magnetic field in spherical coordinates, $\vb B_j = (B_j, \theta_j, \phi_j)$, the spin Hamiltonian \eqref{eq:ZFS-ham} can be decomposed into a longitudinal part,
\begin{equation}
	H_{j,\parallel}^{(e)} = D (S_j^z)^2 + g_e \mu_B B_j \cos(\theta_j) S_j^z ,
\end{equation}
which is diagonal in the eigenbasis $\lbrace \ket{-1}_j, \ket{0}_j, \ket{1}_j \rbrace$ of $S_j^z$, and a transversal part,
\begin{equation}
	H_{j,\perp}^{(e)} = \frac{1}{2} g_e \mu_B B_j \sin(\theta_j) \left( e^{-\rmi \phi_j} S_j^+ + e^{\rmi \phi_j} S_j^- \right) ,
\end{equation}
with ladder operators $S_j^\pm = S_j^x \pm \rmi S_j^y$. In the following we consider a magnetic field intense enough (e.g., $B_j = B = \SI{30}{G}$, for each $j$, such that $g_e \mu_B B = \SI{84}{MHz}$ as in \cite{Bermudez2011}) to ignore the hyperfine coupling (typically a few MHz \cite{Loubser_1978}) between the electron spin and the spin of the nitrogen nucleus.

In the system described by Eq.~\eqref{eq:ZFS-ham}, the transition $\ket{0}_j \leftrightarrow \ket{-1}_j$ is addressed as a qubit, that is, it is driven by an external electromagnetic field, which we here assume to be monochromatic and resonant with the transition frequency $\omega_j = D - g_e \mu_B B \cos\theta_j$. In the dipole approximation \cite{Loudon2010}, the Hamiltonian mediating this driving is
\begin{equation}
	H_j^\mathrm{d} = \Omega \cos(\omega_j t) \sigma_j^x ,
\end{equation}
with $\Omega = \SI{15}{MHz}$ \cite{Bermudez2011} the Rabi frequency 
\cite{Loudon2010}
and $\sigma_j^x = \ket{-1}_j\bra{0}_j + \ket{0}_j\bra{-1}_j$. %

Finally, we add the dipolar coupling between the electronic spins of NV centers $j$ and $k$:
\begin{align}\label{eq:dipolar-coupling}
	H_{jk}^{(ee)} & = J_{jk} \left(\vb S_j \cdot \vb S_k - 3 S_j^z S_k^z\right) \nonumber \\
	& = \frac{J_{jk}}{2} \left( S_j^+ S_k^- + S_j^- S_k^+ - 4 S_j^z S_k^z\right) ,
\end{align}
with $S_j^\pm = S_j^x \pm \rmi S_j^y$, $J_{jk} = \mu_0 \gamma_e^2/4\pi r_{jk}^3 \approx \SI{70}{kHz} $ for a distance $r_{jk} \approx \SI{10}{nm}$ between NV centers $j$ and $k$ \cite{Neumann_2010}, $\gamma_e = g_e \mu_B / \hbar$ the gyromagnetic ratio of the electron, and we have already taken into account that the NV centers are placed on the $z$ axis.

The Hamiltonians above describe dynamics on different timescales, since $\omega_j = D - g_e\mu_B B \cos\theta_j \gg \Omega \gg J_{jk}$, with $H_{j,\parallel}^{(e)}$ describing the fastest and $H_{jk}^{(ee)}$ the slowest dynamics. By moving to the interaction picture with respect to $H_{j,\parallel}^{(e)}$, we can therefore neglect rapidly oscillating terms, whose contribution to the energy averages to zero on the timescale of the slowest dynamics, to simplify the Hamiltonian under study. In this secular approximation, the Hamiltonian of a chain of NV centers reads
\begin{equation}\label{eq:chain_Hamiltonian}
	H = \sum_j \frac{\Omega}{2} \sigma_j^x - \sum_{k>j=1} \frac{J_{jk}}{2} \left(\sigma_j^z - \mathbb{I}\right) \left(\sigma_k^z - \mathbb{I}\right).
\end{equation}
To obtain this spin Hamiltonian we have neglected the transversal part $H_{j,\perp}^{(e)}$ of the electronic Hamiltonian, the counter-rotating terms in $H_j^\text{d}$, and the transversal terms in $H_{jk}^{(ee)}$. The latter terms can be dropped only when an inhomogeneous magnetic field renders the Zeeman splittings of distinct NV centers sufficiently different, $g_e \mu_B B |\cos\theta_j - \cos\theta_k| \gg J_{jk}$, such that the operators $S_j^+ S_k^- $ and $ S_j^- S_k^+$ in Eq.~\eqref{eq:dipolar-coupling} acquire phases rotating fast on the timescale induced by the dipolar coupling $ J_{jk}$.
Furthermore, since we are resonantly driving the transition $\ket{0}_j \leftrightarrow \ket{-1}_j$, we restrict the $S_j^z$ operators remaining in Eq.~\eqref{eq:dipolar-coupling} 
to the subspace spanned by these two (local) states, by writing $S_j^z = \ket{1}_j\bra{1}_j + (\sigma_j^z - \mathbb{I})/2$ and then neglecting the projector on the state $\ket{1}_j$, which we never populate. We thus define $\sigma_j^z = \ket{0}_j\bra{0}_j - \ket{-1}_j\bra{-1}_j$ and $\mathbb{I}$ as the identity operator on this two-dimensional space.

Since $J_{jk} / \Omega \ll 1$, the eigenstates of Eq.~\eqref{eq:chain_Hamiltonian} are distributed in energy manifolds separated by $ \Omega$, in analogy to the spectrum of the unperturbed Hamiltonian [i.e., $ J_{jk} = 0$ in \eqref{eq:chain_Hamiltonian}]. Up to perturbative corrections, therefore, these manifolds can be identified by the number of excitations carried by their states, which is a well-defined quantum number for the unperturbed Hamiltonian. Notably, the lowest- and highest-energy manifolds consist of only one (separable) state each, the perturbative corrections of, respectively, $\ket{-,-,\ldots,-}$ and $\ket{+,+,\ldots,+} $, with $\ket{\pm} = (\ket{0} \pm \ket{1})/\sqrt{2}$ the eigenbasis of $\sigma_j^x$ for each site $j$. 

Moreover, within the same perturbative picture, manifolds with $k$ and $N-k$ excitations have the same structure: exchanging $\ket{0}$ with $\ket{-1}$ leaves $\sigma_j^x$ unchanged, and only flips the sign of $\sigma_j^z$. The latter effect changes the sign of local terms like $\sigma_j^z \otimes \mathbb{I}$, which affect the separation between manifolds, but does not change the basic interaction term $\sigma_j^z \otimes \sigma_k^z$, which determines the entanglement structure of the eigenstates of Eq.~\eqref{eq:chain_Hamiltonian}. In conclusion, it is sufficient to study only the entanglement properties up to the $\lfloor (N+1)/2 \rfloor$-excitation manifold, and
from now on every remark made about the $k$-excitation manifold will also hold for $N-k$ excitations.
We have nevertheless checked numerically that the properties are mirrored in the other half of the spectrum.

\begin{figure}
	\centering
	\includegraphics[width=\columnwidth]{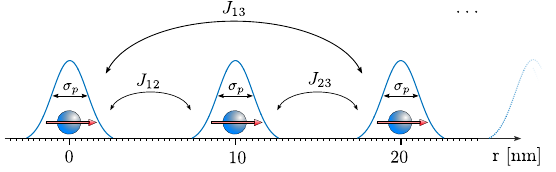}
	\caption{Spin chain model of nitrogen-vacancy centers spatially aligned along a dislocation (a one-dimensional defect in diamond), represented by the axis. For the regularly spaced chain, the spins (spheres, with arrows indicating the aligned symmetry axes of the NV centres, see Sec.~\ref{sec:chains}) are $\SI{10}{\nano\meter}$ apart (major ticks). The minor ticks correspond to the lattice constant of diamond ($\SI{0.3567}{\nm}$), such that the major ticks are separated by 28 minor ticks. When we introduce positional disorder along the dislocation, the new position of each spin is drawn from a Gaussian distribution centered on the original position, with variable width $\sigma_p$ controlling the strength of the disorder. All spins interact with each other via dipolar coupling of strength $J_{jk}$, see Eq.~\eqref{eq:dipolar-coupling}.}
 	\label{fig:spin-chain}
\end{figure}

Our starting point will be the entanglement structures in regularly spaced chains of NV centers, each separated by a distance of approximately $28a$, with $a=\SI{0.3567}{nm}$ \cite{Holloway1991} the lattice constant of the diamond host.
We then model disorder by drawing, for each realization, an array $\{r_j\}_{j = 1, \dots , N}$ for the spin positions, where the position $r_j$ of center $j$ is drawn randomly from a Gaussian distribution centered on the spin's original position in the regularly spaced chain, see Fig.~\ref{fig:spin-chain}. The strength of the disorder is then controlled by the Gaussian's width $\sigma_p \in \{0.1, 0.2, 0.4, 0.8\}\,\unit{nm}$, where the strongest disorder corresponds to a $3\sigma_p$-deviation of almost $7a$ from the original position.

Given the last remark, we have to stress that we are here assuming, for simplicity, that the NV centers are point-like objects that can move continuously on a line.
This assumption is sufficient for the sake of investigating the entanglement structures in the eigenstates, since it mimics the possibility that the NV centers are translated by a few lattice constants, and also that the lattice is itself slightly deformed due to, e.g., strain fields. What matters is that, for these disorder strengths, the spectrum still consists of energetically well-separated manifolds, and, more fundamentally, that the NV centers do not get so close that their electronic structure is unrecognizable, to the point that we cannot start from Eq.~\eqref{eq:ZFS-ham} anymore.

\subsection{Entanglement quantification}\label{sec:ent-quantification}
The \emph{entropy of entanglement} \cite{Nielsen2010} is a quantifier for pure states that determines how strongly a subset $A \subset \lbrace 1, 2, \ldots, N \rbrace$ of spins is entangled with the rest of the chain, denoted by the complementary set $\bar{A}$. Given the non-vanishing eigenvalues $\lbrace \lambda_j\rbrace_j$ of the reduced state $\rho_A = \tr_{\bar{A}}\ket{\Psi}\bra{\Psi}$, obtained by the partial trace over the complementary degrees of freedom, the entropy of entanglement reads
\begin{equation}\label{eq:ent-entropy}
	E_A(\ket{\Psi}) = - \sum_j \lambda_j \log_2 \lambda_j .
\end{equation}
If $\ket{\Psi}$ separates along the bipartition $\lbrace A, \bar{A} \rbrace$, i.e., $\ket{\Psi} = \ket{\psi}_A \ket{\phi}_{\bar{A}}$, then $\rho_A$ is pure, all of its eigenvalues vanish except for one, and therefore $E_A(\ket{\Psi}) = 0$. In the following, whenever we find a bipartition where the entropy of entanglement vanishes, we will conclude that the probed state $\ket{\Psi}$ separates along the given bipartition. If, on the contrary, no bipartition exists where the entropy of entanglement vanishes, we call the state $N$-partite entangled.

The other quantifier we will consider in the following is \emph{Wootters's concurrence} \cite{Wootters1998, PhysRevA.61.052306} of a mixed state $\rho_A$ of two spins $j$ and $k$, defined, as above, as $\rho_A = \tr_{\bar{A}}\ket{\Psi}\bra{\Psi}$ with $A = \lbrace j,k\rbrace$. Wootters's concurrence therefore reads
\begin{equation}\label{eq:concurrence}
	C_{jk}(\ket{\Psi}) = \max( 0, \mu_1 - \mu_2 - \mu_3 - \mu_4) ,
\end{equation}
where $\mu_1, \ldots, \mu_4$ are the eigenvalues, in decreasing order, of the matrix $\sqrt{\sqrt{\rho_A}\tilde{\rho}_A\sqrt{\rho_A}}$, with $ \tilde{\rho}_A = (\sigma_y \otimes \sigma_y)\rho_A^*(\sigma_y \otimes \sigma_y)$, and ${}^*$ the element-wise complex conjugation.
It can be shown that this quantity attains its maximum (unity) for the Bell states \cite{PhysRevA.64.042315}. %

The concurrence quantifies the entanglement left in the joint state of spins $j$ and $k$ after tracing away the degrees of freedom of all other spins. With the latter operation, however, we necessarily discard any entanglement that connects the selected spins to the others.
For this reason, we have to complement the concurrence between each spin pair with the entropy of entanglement as defined in Eq.~\eqref{eq:ent-entropy}.

A typical example of an entangled state is the W state \cite{Duer_2000}, which, for $N$ qubits, is written as
\begin{equation}\label{eq:W-state}
	\ket{W_N} = \frac{\ket{10\dots0}+\ket{01\dots0}+\dots+\ket{0\dots01}}{\sqrt{N}} .
\end{equation}
Notice that we have now switched to the \emph{computational basis} \cite{Nielsen2010}, with $\lbrace \ket{0}, \ket{1} \rbrace$ the new labels for the states $\lbrace \ket{0}, \ket{-1} \rbrace$ used in Sec.~\ref{sec:ent-quantification}.
As we show in App.~\ref{sec:app-proofs}, these states are 
$N$-partite entangled,
\begin{equation}\label{eq:W-entropy}
	E_A(\ket{W_N}) = - \frac{N-\ell}{N} \log_2\left(\frac{N-\ell}{N}\right) - \frac{\ell}{N} \log_2\left(\frac{\ell}{N}\right),
\end{equation}
with $\ell$ the number of spins included in $A$, and
\begin{equation}\label{eq:W-concurrence}
	C_{jk}(\ket{W_N}) = \frac{2}{N}, \qquad \forall j,k .
\end{equation}

Concomitant with the W state we also find the GHZ state \cite{Greenberger_1993}, defined for $N \geq 3$ qubits as
\begin{equation}\label{eq:GHZ-state}
	\ket{GHZ_N} = \frac{\ket{00\ldots 0}+\ket{11\ldots1}}{\sqrt{2}},
\end{equation}
and characterized by uniform entanglement entropy across all bipartitions,
\begin{equation}\label{eq:GHZ-entropy}
	E_A(\ket{GHZ_N}) = 1, \qquad \forall A,
\end{equation}
as well as vanishing pairwise concurrences,
\begin{equation}\label{eq:GHZ-concurrence}
	C_{jk}(\ket{GHZ_N}) = 0, \qquad \forall j,k .
\end{equation}
Remarkably, for the GHZ state all pairwise concurrences vanish, and yet the state is $N$-partite entangled.

Let us comment here about the \emph{structure} of these states. If we identify the elements of the computational basis with the bit strings (the sequences of `0' and `1') that identify them, we notice that the W state is the balanced sum of all strings where all bits except one take the same value, while the GHZ state is the sum of a string and its bit-wise negation (or complement).
From these observations, we can define \emph{generalized} W states as
	\begin{equation} \label{eq:gen-W}
		\ket{W^\text{gen.}_N}  = 
		c_1\ket{10\dots0}+c_2\ket{01\dots0}+\dots+c_N\ket{0\dots01} ,
	\end{equation}
with $c_j$'s non-vanishing complex coefficients (satisfying normalization). The properties of this state, which we will use later to discriminate among entanglement structures, are
\begin{equation} \label{eq:W-like}
	\begin{aligned}
		& E_A(\ket{W^\text{gen.}_N}) \neq 0, \quad \forall A, \\
		& C_{jk}(\ket{W^\text{gen.}_N}) = 2|c_jc_k| \neq 0, \qquad \forall j,k ,
	\end{aligned}
\end{equation}
where the concurrence is correctly bounded by unity when $|c_jc_k| \leq 1/2$.
	Furthermore, for any orthonormal basis $\lbrace \ket{\phi_j}, \ket{\psi_j} \rbrace$ of the Hilbert space of spin $j$, we can define \emph{generalized} GHZ states \cite{Minke:2023aa} as
\begin{equation}
	\ket{GHZ^\text{gen.}_N}  = c_1 \ket{\phi_1\ldots \phi_N} + c_2 \ket{\psi_1\ldots\psi_N} \label{eq:gen-GHZ},
\end{equation}
where again $c_1$ and $c_2$ are non-vanishing complex coefficients satisfying $|c_1|^2+|c_2|^2=1$. The properties of these states are
 \begin{equation}\label{eq:GHZ-like}
 	\begin{aligned}
 		 	& E_A(\ket{GHZ^\text{gen.}_N}) \neq 0, \quad \forall A, \\
 			& C_{jk}(\ket{GHZ^\text{gen.}_N}) = 0, \quad \forall j,k .
 	\end{aligned}
 \end{equation}
In particular, according to Eq.~\eqref{eq:gen-GHZ}, $(\ket{1001} - \ket{0110})/\sqrt{2}$ is also a GHZ state.

\begin{figure*}
	\centering
	\includegraphics[width=\textwidth]{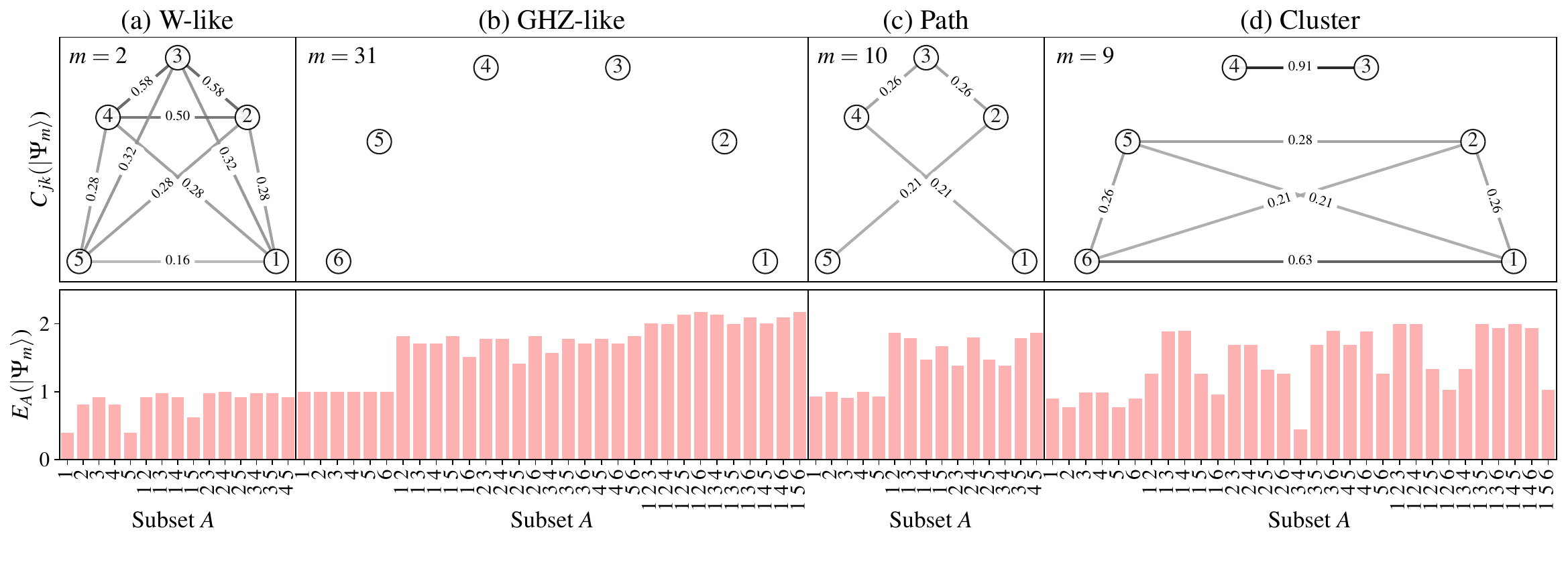}
	\caption{Entanglement structures (columns) of exemplary eigenstates $\ket{\Psi_m}$ of regularly spaced chains of 5 [(a) and (c)] or 6 [(b) and (d)] NV centers.
		The \emph{top row} shows graphs with the NV centers as nodes and pairwise concurrences $C_{jk}(\ket{\Psi_m})$ [Eq.~\eqref{eq:concurrence}] as edges. The opacity of the edges is proportional to the concurrence, whose value is also indicated, and whenever $C_{jk}(\ket{\Psi_m}) \leq 0.005$ the edge is not shown. For compactness, nodes have been placed on a (suitably stretched) semicircle, even though the NV centers form a chain. The bar plots in the \emph{bottom row} show the entropy of entanglement $E_A(\ket{\Psi_m})$ [Eq.~\eqref{eq:ent-entropy}]  for the bipartition $\lbrace A, \bar{A} \rbrace$ stemming from the subset $A$ of NV centers indicated on the abscissae.
		Note that $E_A(\ket{\Psi_m})$ is bounded between 0 and the cardinality $|A|$ of the subset $A$.
	}%
\label{fig:ent-structure-regular}
\end{figure*}

\subsection{Automated classification of entanglement structures}\label{sec:automated-classification}
The aim of our analysis is to establish, for each eigenstate of \eqref{eq:chain_Hamiltonian}, which spins in the chain are entangled with each other, and how, by combining the information on the entropy of entanglement and on the concurrence.
To this end, we define entropies as non-vanishing whenever they exceed the threshold $\epsilon = 0.01$.
This threshold is necessary because the algebraic tail of the dipolar coupling ($J_{jk} \propto r_{jk}^{-3}$) formally couples \emph{all} spins in the chain \cite{Anderson:1958aa}. We therefore need a criterion to distinguish strongly from weakly correlated states, which can be assimilated to truly separable (uncoupled) states.
Given this definition, we classify the entanglement structure of the eigenstate $\ket{\Psi_m}$ as follows:
\begin{itemize}
	\item \emph{W-like}: 
	if $E_A(\ket{\Psi_m}) \neq 0, \forall A$ and $C_{jk} (\ket{\Psi_m}) \neq 0, \forall j,k$.
	Plotted as a graph, edges connect each node (the spins) to all the others, as shown in Fig.~\ref{fig:ent-structure-regular} (a).
	These are states with the same entanglement structure [cf.~Eq.~\eqref{eq:W-like}] as the generalized W state in Eq.~\eqref{eq:gen-W}.
	\item \emph{GHZ-like}: if $E_A(\ket{\Psi_m}) \neq 0, \forall A$ and $C_{jk} (\ket{\Psi_m}) = 0, \forall j,k$.
	 Here, no edge is visible in the graph, see Fig.~\ref{fig:ent-structure-regular} (b). As explained in Sec.~\ref{sec:ent-quantification}, this state is still $N$-partite entangled.
	 These are states with the same entanglement structure [cf.~Eq.~\eqref{eq:GHZ-like}] as the generalized GHZ state in Eq.~\eqref{eq:gen-GHZ}.
	\item \emph{Path}: if $E_A(\ket{\Psi_m}) \neq 0, \forall A$ and $C_{jk} (\ket{\Psi_m}) = 0$ for $j$ and $k$ such that the edges of non-zero concurrence form a path that visits all the nodes (not necessarily in order), as in Fig.~\ref{fig:ent-structure-regular} (c).
	\item \emph{Cluster}: if $E_A(\ket{\Psi_m}) \neq 0, \forall A$ and $C_{jk} (\ket{\Psi_m}) = 0$ for $j$ and $k$ such that the concurrence graph splits in two or more subgraphs. For the exemplary state in Fig.~\ref{fig:ent-structure-regular} (d), spins in the subset $A = \lbrace 3,4 \rbrace$ form a disjoint subgraph from the other spins. Since $E_{\lbrace 3,4\rbrace}(\ket{\Psi_m}) < E_{A}(\ket{\Psi_m})$, $\forall A \neq \lbrace 3,4 \rbrace $,
	we conclude that the spins 3 and 4 are only weakly entangled with the rest of the system, although the state remains $N$-partite entangled.
	Let us also stress that `cluster' here describes how the concurrence graph splits, and is not referring to cluster states in two-dimensional spin lattices \cite{Briegel_2001}.
	\item \emph{(Partly) separable}: if $E_A(\ket{\Psi_m}) = 0$ for at least one subset $A$, as shown in Fig.~\ref{fig:ent-structure-separable}.
	Even though the graph in Fig.~\ref{fig:ent-structure-separable} (a) does not formally disjoin, due to very small residual pairwise concurrences, we still consider this state as partly separable, since the entanglement entropy for the spins $\lbrace 2,4\rbrace$ effectively vanishes [i.e., $E_{A=\lbrace 2,4\rbrace}(\ket{\Psi_m}) < \epsilon = 0.01$]. In this case, the spins individually separate from the others, which themselves form a state close to $\ket{W_3}$.
\end{itemize}

In those cases where we need to investigate more closely the structure of the eigenstate $\ket{\Psi_m}$, we will expand it either in the computational basis, or in the eigenbasis $\ket{\pm} = (\ket{0} \pm \ket{1})/\sqrt{2}$ of $\sigma_j^x$ for each qubit $j$. This latter basis is usually most convenient, since it is the eigenbasis of the unperturbed Hamiltonian [i.e., $ J_{jk} = 0$ in \eqref{eq:chain_Hamiltonian}, see Sec.~\ref{sec:chains}]. Such a change in local basis does not affect the entanglement properties of the eigenstates, as already observed in Sec.~\ref{sec:ent-quantification} above.

\section{Results and discussion}\label{sec:discussion}
\begin{figure*}
	\centering
	\includegraphics[width=\linewidth]{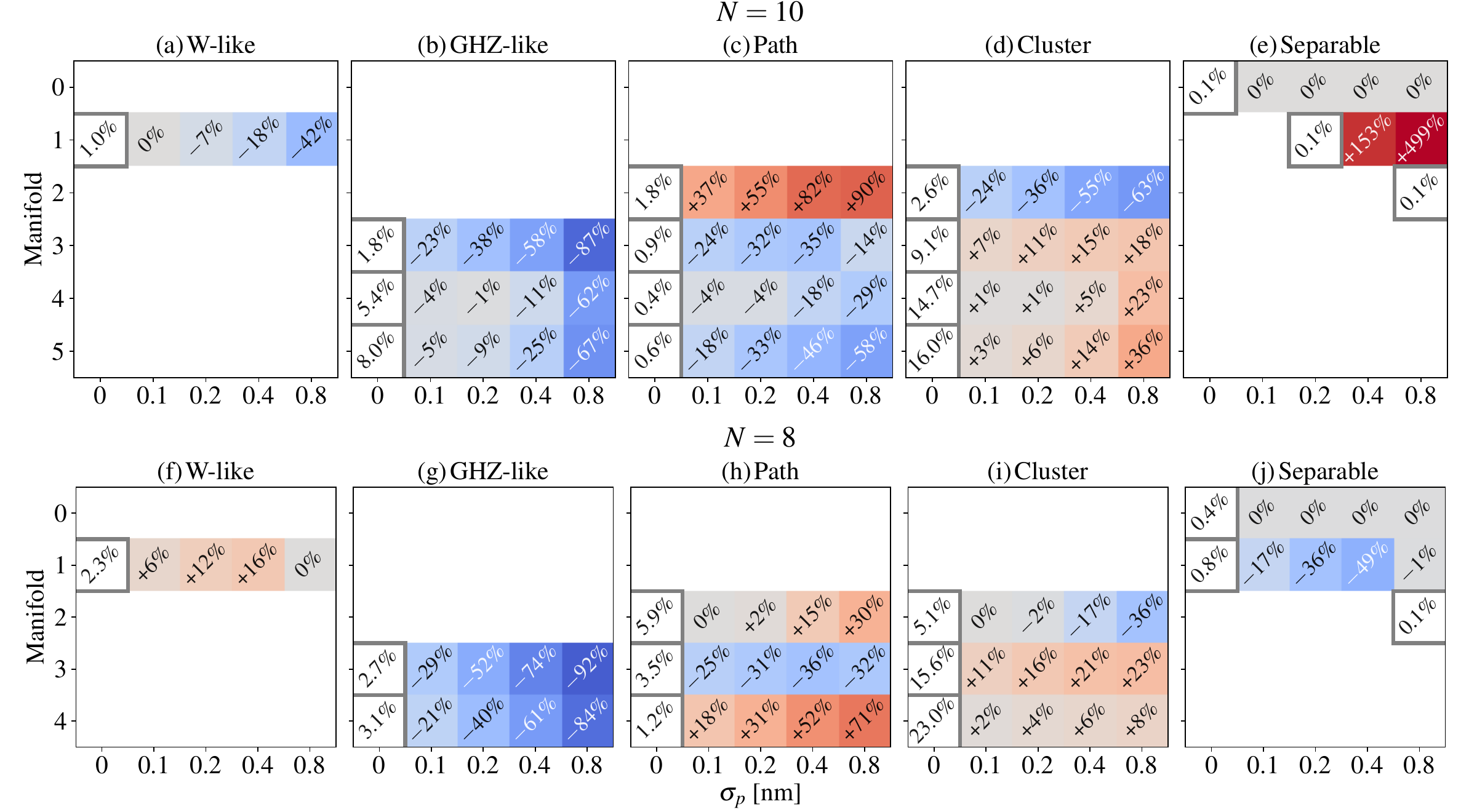}
	\caption{
		Average occurrence frequency of the different entanglement structures defined in Sec.~\ref{sec:automated-classification} (columns).
		The upper row shows the average frequencies in disordered chains of $N = 10$ spins, while the lower row for $N = 8$ spins.
		Each panel is a matrix reporting the occurrence frequencies for manifolds with increasing number of excitations (along its columns), and for increasing disorder strength $\sigma_p$ (along the rows).
		Note that $\sigma_p = \SI{0}{nm}$ indicates a regularly spaced chain.
		To ease legibility, we only report frequencies larger than $0.05\%$. Furthermore, to highlight the trends, the leftmost non-vanishing value in each row [essentially the value for $\sigma_p = \SI{0}{nm}$, except for panels (e) and (j)] is always our \emph{baseline}, the occurrence frequency of the entanglement structure in a given manifold at $\sigma_p$ (reported as percent of the whole spectrum), while the values to its right are the variation along the row compared to this baseline (and therefore may take values larger than 100\%). Variations are indicated both textually and graphically with a temperature map: the baseline of each row is always in a box with a gray border, values larger than the baseline take an increasingly darker red background (and a `+' prefix), while smaller values take an increasingly bluer background (and a `$-$'). At a glance we can therefore read which structures occur more or less frequently with disorder, and in which manifold.}
	\label{fig:ent-structure-10qb}
\end{figure*}

As stated in the Introduction, we prize states where the chain is fully connected, i.e., where all spin pairs are entangled, or all spins participate in an $N$-partite entangled state \cite{Bradley_2019}.
Our first goal is therefore to determine how frequently the entanglement structures from Sec.~\ref{sec:automated-classification} appear in the spectra of regularly spaced chains of $N = 4, \ldots, 10$ spins (reported in Fig.~\ref{fig:ent-structure-regular}).
We subsequently generate $10^3$ realizations of random chains for each system size and disorder strength (cf.\ Sec.~\ref{sec:chains}), and run the classification algorithm of Sec.~\ref{sec:automated-classification} for each obtained spectrum. Since it is not possible to continuously monitor how individual eigenstates change with disorder, we determine the frequency of occurrence of each entanglement class in each manifold (see Fig.~\ref{fig:ent-structure-10qb}), since the latter remain energetically well separated for the disorder strengths considered above.
In this way we can establish how likely it is to generate $N$-partite entangled states, and with them fully connected quantum registers.

\begin{figure*}
	\centering
	\includegraphics[width=\textwidth]{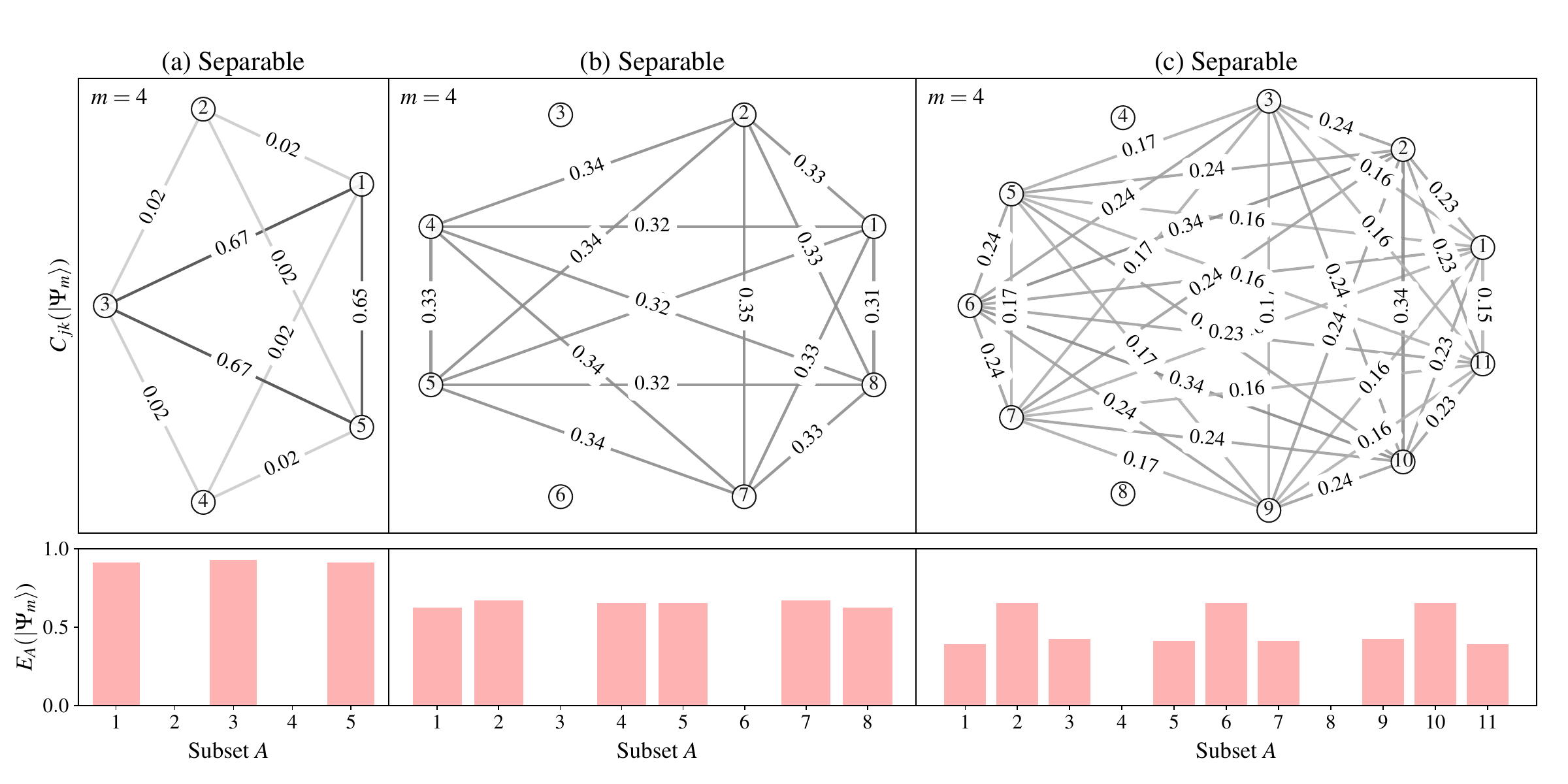}
	\caption{Separable eigenstates of the $1$-excitation manifold of regularly spaced chains of (a) $N=5$, (b) 8, and (c) 11 NV centers. The structures are displayed analogously to Fig.~\ref{fig:ent-structure-regular}, except that the plots in the lower row only show the entropy of entanglement for bipartitions stemming from single-qubit subsets $A$.
		}
	\label{fig:ent-structure-separable}
\end{figure*}

\paragraph{W-like states}

\begin{figure}
	\centering
	\includegraphics[width=\linewidth]{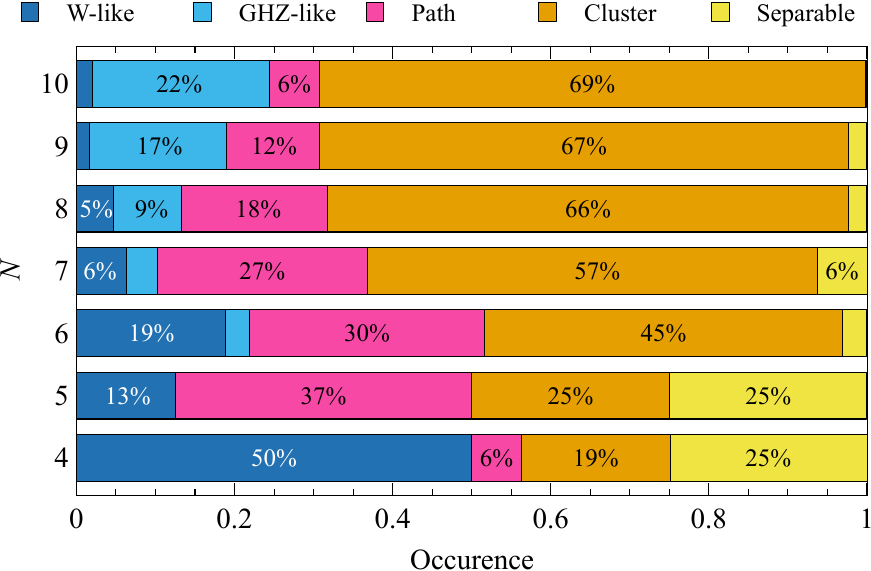}
	\caption{Occurrence of entanglement structures defined in Sec.~\ref{sec:automated-classification} (legend), in the eigenstates of regularly spaced chains of length $N=4,\ldots,10$. 
		If $M$ eigenstates in the whole spectrum of a chain of $N$ spins show a specific entanglement structure, the resulting occurrence will be $M/2^N$.		
		For reasons of space, occurrences smaller than $5\%$ are not reported. This is the case, e.g., for the 20 W-like states in a $N=10$ chain, which have an occurrence of $\approx 2\%$.}
	
	\label{fig:occurrences-ordered}
\end{figure}

These states belong to the 1-excitation manifold, as can be deduced from the definition of the W state in Eq.~\eqref{eq:W-state}.
Whether the 1-excitation manifold consists exclusively of W-like states, depends actually on $N$.
When $N \in \{4,6,10\}$,
all $2N$ states of this manifold are W-like entangled, cf.\ Fig.~\ref{fig:occurrences-ordered}.
In the presence of increasingly stronger disorder,
these states become (partly) separable,
as can be appreciated by comparing panels (a) and (e) in Fig.~\ref{fig:ent-structure-10qb} for an exemplary chain of $N=10$ spins.
The relative abundance of W-like states that become separable depends furthermore on the system size.

For odd lengths and $N = 8$, however, some states are already partly separable in the regularly spaced chain. %
What normally happens for odd $ N $ is that the central spin separates, i.e., some eigenstates become separable along a bipartition that divides the central spin from the others. However, for $N = 5$ and $N = 8$, an additional pattern appears, shown in Fig.~\ref{fig:ent-structure-separable} (a) and (b), respectively, where two opposite-facing spins (with respect to the chain center) separate.
Beyond the central symmetry that all eigenstates have to respect, what these two patterns have in common is that each separating spin is surrounded by an equal number of entangled spins on each side. 
By applying this rule, we found another state with two separable spins in a chain of $N=11$ [Fig.~\ref{fig:ent-structure-separable} (c)].
Curiously, the entangled spins in Fig.~\ref{fig:ent-structure-separable} (a, b) are described by a (true) W state, as in Eq.~\eqref{eq:W-state} (cf. \footnote{In the local eigenbasis $\ket{\pm}$ of $\sigma_j^x$ for each qubit $j$, and up to fluctuations of order $10^{-2}$ in the amplitudes of each basis state.}), and thus fulfill Eq.~\eqref{eq:W-concurrence}. The entangled spins in Fig.~\ref{fig:ent-structure-separable} (c), instead, show a more general W-like structure, rather like Eq.~\eqref{eq:gen-W} in Sec.~\ref{sec:automated-classification}.

These eigenstates become separable because specific amplitudes vanish in the chosen basis. Imagine setting one of the $c_k = 0$ in Eq.~\eqref{eq:gen-W}:
this renders the state separable in spin $k$. In the states of Fig.~\ref{fig:ent-structure-separable}, the amplitudes of the specific basis states [in Eq.~\eqref{eq:gen-W}] where the separable spins are excited \footnote{That is, $\ket{-+---}$ and $\ket{---+-}$ in panel (a), $\ket{--+-\cdots}$ and $\ket{\cdots-+--}$ in panel (b), and so on.} vanish because of the regular spacing between the spins, in what one could consider a (destructive) interference effect.
This intuition is confirmed by turning on positional disorder: by breaking the regularity of the chain we also suppress the interference.
The consequence is that the occurrence of W-like entanglement rises for $\SI{0.1}{nm} \leq \sigma_p \leq \SI{0.4}{nm}$, as shown for the 
$N=8$ chain in Fig.~\ref{fig:ent-structure-10qb}\,(f) and (j).
Still, when disorder becomes sufficiently strong ($\sigma_p=\SI{0.8}{nm}$), W-like entanglement is lost to separability.
A deviation from this trend are chains of length $N=5$, where the destruction of W-like entanglement requires stronger disorder (cf.~$N=5,7,9$ in Fig.~\ref{fig:all_data_disorder}).

A similar pattern appears for $N=4$ in Fig.~\ref{fig:all_data_disorder}. For that system size only, there are two states in the 2-excitation manifold that are separable (along $A=\lbrace1,4\rbrace$), also because of destructive interference of amplitudes of specific basis states. If we increase the degree of disorder, in some ensemble these states will become path or cluster states, although, in a few cases, they will take a W-like structure. This explains why we see an increase in the occurrence of this structure for $\sigma_p > 0$ for $N=4$. This is the only system size where we have seen W-like structures appear outside of the 1-excitation manifold.

In summary, W-like states, where all qubits are pairwise connected by bipartite entanglement, are typical of the 1-excitation manifold, and appear most frequently in the presence of weak disorder, which avoids separability due to destructive interference.

\paragraph{GHZ-like states}

Figure \ref{fig:occurrences-ordered} reveals that GHZ-like states appear for $N \geq 6$ and occur more frequently in larger chains. Moreover, as Fig.~\ref{fig:ent-structure-10qb} shows, these states appear starting from the 3-excitation manifold, and with increasing occurence when moving closer to the center of the spectrum (the $N/2$- and $(N - 1)/2$-excitation manifold for even and odd $N$, respectively). Recall that a true GHZ state [Eq.~\eqref{eq:GHZ-state}] cannot appear in the spectrum of the Hamiltonian, which consists of manifolds where the number of excitations is well defined (up to perturbative corrections). What we observe here are eigenstates with the same entanglement structure as the generalized GHZ states in Eq.~\eqref{eq:gen-GHZ}, yet consist of more than two basis states (of either the computational basis or the eigenbasis of $\sigma_j^x$). It remains unclear whether local rotations can bring these GHZ-like eigenstates to the form of Eq.~\eqref{eq:gen-GHZ}.

The introduction of disorder leads to a (nearly) monotonous decrease in the occurrence of GHZ-like states for all chain lengths, with a larger relative loss in the energetically lower manifolds, as can be read off from Fig.~\ref{fig:ent-structure-10qb}\,(b) and (g).
Similarly to the destructive interference discussed above, the loss of symmetry in the chain changes the state (e.g., in its decomposition in a chosen basis), such that the vanishing concurrences, which define GHZ-like states, increase to small yet non-vanishing values, whereas the entropies of entanglement never vanish, implying $N$-partite entanglement. For this reason, with increasing disorder GHZ-like transform into cluster states.

In short, even though GHZ-like states become more frequent with increasing system size, they are susceptible to noise, and are likely to transform into cluster states.

\paragraph{Path states}

These states are found in all manifolds with at least two excitations, and the occurrence of these states in the spectrum reaches its maximum in a 5-spin chain (see Fig.~\ref{fig:occurrences-ordered}), and then decreases again for larger systems.
As for their distribution across manifolds, the trend in Fig.~\ref{fig:ent-structure-10qb}\,(c) is exemplary:
the $2$-excitation manifold always contains the largest share of path states, and this share decreases towards the center of the spectrum.

With exceptions that will be discussed immediately below, the response of path states to disorder follows general patterns: 
in the 2-excitation manifold the share of this entanglement structure increases, whereas, starting from the 3-excitation manifold, it tends to decrease (not necessarily monotonously) with disorder strength.
The most glaring exception to this pattern is the increase with disorder of the abundance of path states within the 4-excitation manifold of the 8-spin chain, see Fig.~\ref{fig:ent-structure-10qb}\,(h). 
Here two possibilities arise: either GHZ-like entangled states acquire path instead of cluster entanglement, or GHZ-like become cluster entangled states, and formerly cluster transform into path entangled states. Contrarily to the discussion for W-like states, we cannot ascertain here whether interference effects are at play.
We otherwise never see an increase in the occurrence of path states outside of the 2-excitation manifold. Even there, for instance
for $N=4$ and $5$, 
this increase is not monotonous, but rather shows a local minimum near $\sigma_p = \SI{0.2}{nm}$.

All in all, path structures become increasingly untypical for larger registers, starting from $N \geq 7$. This observation makes their robustness to disorder less relevant.

\paragraph{Cluster states}
Analogously to path states, cluster states are found in manifolds with at least two excitations, yet show completely opposite trends compared to the previous entanglement structure. The occurrence of cluster states in the spectrum increases monotonously with the size of the system (compare Fig.~\ref{fig:occurrences-ordered}), and
the majority of cluster-entangled states comes from the $3$-excitation manifold and above, where cluster states make up approximately $60\%$ to $80\%$ of the manifolds \footnote{For instance, for $N=10$, we read in Fig.~\ref{fig:ent-structure-10qb} (d) that cluster states in the 5-excitation manifold make up $16\%$ of the spectrum, that is $0.16 \times 2^{10}\approx 164$ states, as compared to all $\binom{10}{5} = 252$ states in the chosen manifold.}.

Contrarily to path states, when we increase disorder the occurrence of cluster states in the 2-excitation manifold reduces, while it typically increases in the manifolds closer to the center of the spectrum.
It is worth noting that the 2-excitation manifold consists, almost exclusively, of path and cluster states. This means that any change of the abundance of one class is compensated by that of the other, since the number of states in the manifold is conserved. 
This reasoning does not hold from the $3$-excitation manifold onward, since GHZ-like states can also transform into either path or cluster states.
Furthermore, for $\sigma_p = \SI{0.8}{nm}$ some states in the 2-excitation manifold lose their entanglement properties, as few separable states can be seen in Fig.~\ref{fig:ent-structure-10qb}\,(e) and (j), as well as in Fig.~\ref{fig:all_data_disorder} for $N=4,7,9$.

Ultimately, the popularity of cluster states in the spectra of ordered and disordered chains of NV centers reflects the fact that they are defined by less stringent requirements than the other classes. While the number of pairwise connections -- via the concurrence -- between qubits might vary, these states are still $N$-partite entangled, as captured by the entropy of entanglement.

\section{Conclusions and outlook}\label{sec:conclusions}
In the present manuscript we have investigated the possibility of creating connected registers on chains of dipole-coupled NV centers. This connectivity is assessed by computing the multipartite and bipartite entanglement inscribed in the eigenstates of the chain, as mediated by the entropy of entanglement and Wootters's concurrence, respectively.
In general, the vast majority of the states is $N$-partite entangled, i.e., no bipartition can be found where the state is separable. More specifically, starting from $N \geq 6$, 
the occurrence of cluster and GHZ-like entanglement progressively increases, while that of W-like and path structures decrease.
In particular, cluster and GHZ-like states
are encountered in the manifolds closer to the center of the spectrum,
which also have the largest dimensions \footnote{The $k$-excitation manifold is spanned by $\binom{N}{k}$ basis states (up to perturbative corrections).}. Generic, delocalized states in those manifolds will therefore naturally be multipartite entangled \cite{Beugeling2015}.

The introduction of positional disorder does not significantly affect $N$-partite entanglement: up to a disorder strength of $\sigma_p=\SI{0.8}{nm}$, corresponding to a couple of lattice constants $a = \SI{0.3567}{\nm}$, only few states turn partly separable. We otherwise observe a general increase in cluster entangled states, at the expense of W-like, GHZ-like and path structures. A notable exception are systems of size $N = 3L+2$, with integer $L \geq 1$, where the 1-excitation manifold contains several separable eigenstates. These form because of the regular spacing of the chain, which makes some of the amplitudes in Eq.~\eqref{eq:gen-W} to vanish by destructive interference. As soon as disorder breaks the regular spacing, thereby suppressing this interference effect, separable states acquire W-like entanglement.

All in all, chains of NV centers assembling along a dislocation can be used to build registers connected by multipartite entanglement, and these registers are resilient against, or even benefit from, relatively weak disorder ($\sigma_p \lesssim \SI{0.4}{nm}$), i.e., from a slight displacement of roughly up to a lattice constant $a$ from the lattice position. Stronger disorder ($\sigma_p \gtrsim \SI{1.6}{nm}$) will eventually lead to the manifolds overlapping, a situation that necessitates a more sophisticated analysis of how the entanglement structures transform into each other.

\begin{acknowledgments}
The authors thank Reyhaneh Ghassemizadeh, Daniel Urban and Tommaso Faleo for insightful exchanges.
The authors acknowledge support by the state of Baden-W\"urttemberg through bwHPC.
E.~G.~C.\ acknowledges support from the Georg H.~Endress foundation, and from the project ``SiQuRe'' (Kompetenzzentrum Quantencomputing Baden-W\"urttemberg) funded by the Ministerium f\"ur Wirtschaft, Arbeit und Tourismus of the State of Baden-W\"urttemberg.
\end{acknowledgments}
\appendix
\section{Entanglement entropy and concurrence of W states}\label{sec:app-proofs}
Let us consider a W state of $N$ two-level systems (qubits), as in Eq.~\eqref{eq:W-state}, and let us consider a subset $A$ containing the first $\ell$ qubits (the other possible subsets we do not need to consider, since the W state is symmetric under permutations of the qubit indices). We can then write \cite{Minke:2023aa} the $\ket{W_N}$ state as
\begin{align}
	\ket{W_N} = & \frac{1}{\sqrt{N}} \left[ \sum_{\pi \in \Sigma_\ell} \ket{\pi(10\ldots0)}_A \otimes \ket{0\ldots0}_{\bar{A}} \right. \nonumber \\ & \left. + \ket{0\ldots0}_A \otimes \sum_{\pi \in \Sigma_{N-\ell}} \ket{\pi(10\ldots 0)}_{\bar{A}} \right] \; ,
\end{align}
where $\Sigma_\ell$ is the set of \emph{distinct} permutations of the $\ell$-letter word `$10\ldots0$', and has therefore cardinality $\ell$. For clarity we have indicated the subset $A$ or $\bar{A}$ to which the states pertain. If we define the normalized states
\begin{align*}
	& \Ket{\chi_1^{(\ell)}} = \Ket{00\ldots 0}_A, \nonumber \\
	& \Ket{\chi_2^{(\ell)}} = \frac{1}{\sqrt{\ell}}  \sum_{\pi \in \Sigma_\ell} \Ket{\pi(10\ldots0)}_A ,
\end{align*}
for subset $A$, and
\begin{align*}
	& \Ket{\phi_1^{(\ell)}} = \frac{1}{\sqrt{N-\ell}} \sum_{\pi \in \Sigma_{N-\ell}} \Ket{\pi(10\ldots 0)}_{\bar{A}}, \nonumber \\
	& \Ket{\phi_2^{(\ell)}} = \Ket{00\ldots 0}_{\bar{A}} ,
\end{align*}
for the complementary set $\bar{A}$, we can  write the Schmidt decomposition
\begin{equation}\label{eq:W-schmidt}
	\Ket{W_N} = \sqrt{\frac{N-\ell}{N}} \Ket{\chi_1^{(\ell)}}\Ket{\phi_1^{(\ell)}} + \sqrt{\frac{\ell}{N}} \Ket{\chi_2^{(\ell)}}\Ket{\phi_2^{(\ell)}} .
\end{equation}
From the Schmidt coefficients $\sqrt{\lambda_1} =\sqrt{(N-\ell)/N} $ and $\sqrt{\lambda_2} = \sqrt{\ell/N}$ (written in decreasing order, if we assume $\ell < N/2$) we can obtain, by direct substitution in Eq.~\eqref{eq:ent-entropy}, the entanglement entropy in Eq.~\eqref{eq:W-entropy}.

To prove Eq.~\eqref{eq:W-concurrence}, instead, we just need to consider partitions $A$ of two qubits. We then have, from \eqref{eq:W-schmidt},
\begin{equation}
	\Ket{W_N} = \sqrt{\frac{N-2}{N}} \Ket{\chi_1^{(2)}}\Ket{\phi_1^{(2)}} + \sqrt{\frac{2}{N}} \Ket{\chi_2^{(2)}}\Ket{\phi_2^{(2)}} .
\end{equation}
The reduced state of the selected qubits will then be
\begin{align}
	\rho_A & = \tr_{\bar{A}} \left[\Ket{W_N}\Bra{W_N}\right] \nonumber \\
	& = \frac{N-2}{N} \Ket{\chi_1^{(2)}}\Bra{\chi_1^{(2)}} + \frac{2}{N} \Ket{\chi_2^{(2)}}\Bra{\chi_2^{(2)}} .
\end{align}
In the basis $\ket{0} \equiv \binom{1}{0}$ and $\ket{1} \equiv \binom{0}{1}$, this density operator is represented by
\begin{equation}
	\rho_A \equiv \frac{1}{N} \begin{pmatrix}
		N-2 & 0 & 0 & 0 \\
		0 & 1 & 1 & 0 \\
		0 & 1 & 1 & 0 \\
		0 & 0 & 0 & 0
	\end{pmatrix} .
\end{equation}
Direct substitution in the procedure described in Sec.~\ref{sec:ent-quantification} finally yields the result in Eq.~\eqref{eq:W-concurrence}.

A completely analogous procedure can be used to derive $C_{jk}(\ket{W^\text{gen.}_N})$ in Eq.~\eqref{eq:W-like}. In this case, the reduced density operator $\rho_A$, when $A=\lbrace j,k \rbrace$, reads
\begin{equation}
	\rho_A \equiv \begin{pmatrix}
		1-|c_j|^2-|c_k|^2 & 0 & 0 & 0 \\
		0 & |c_j|^2 & c_jc_k^* & 0 \\
		0 & c_j^*c_k & |c_k|^2 & 0 \\
		0 & 0 & 0 & 0
	\end{pmatrix} .
\end{equation}

\section{Occurrence of entanglement structures for all chain lengths}
\begin{figure*}
	\centering
	\includegraphics[width=\linewidth]{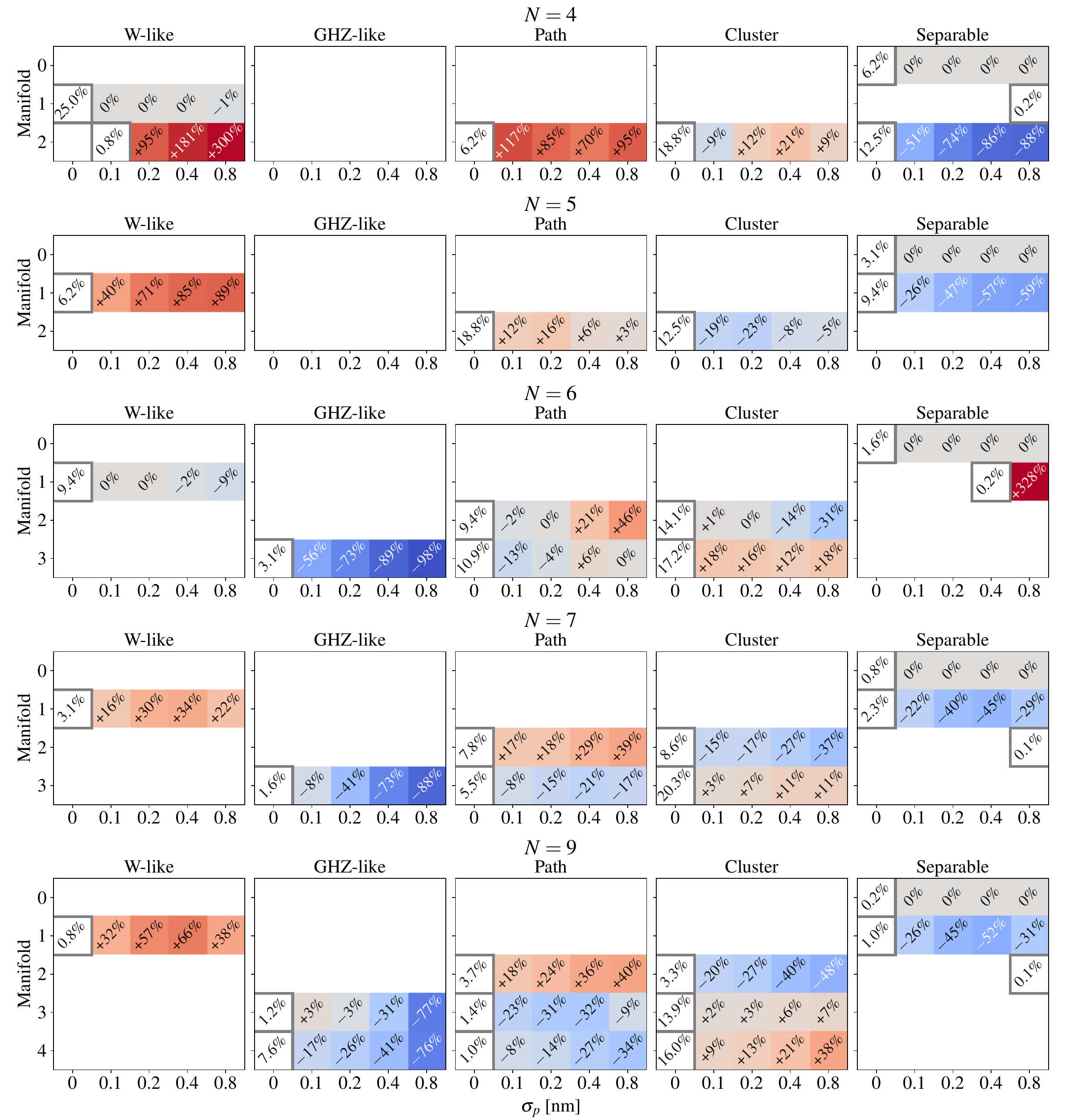}
	\caption{Average occurrence frequency of the different entanglement structures defined in Sec.~\ref{sec:automated-classification} (columns), as a function of disorder strength (rows) for those system sizes here analyzed but not shown in Fig.~\ref{fig:ent-structure-10qb}. See the caption of this latter figure for a description of the organization and visualization of the data.}
	\label{fig:all_data_disorder}
\end{figure*}

In Fig.~\ref{fig:all_data_disorder} we show the change in entanglement structures that can be encountered in the manifolds of spin chains of length $N=4, \ldots, 10$, except for $N=8$ and $10$, which are already shown in Fig.~\ref{fig:ent-structure-10qb}.

%

\end{document}